\documentclass[twocolumn]{aastex631}

\newcommand{\angstrom}{\textup{\AA} }

\newcommand{\nev}{\mbox{[Ne \textsc{v}] }}
\newcommand{\nevns}{\mbox{[Ne \textsc{v}]}}
\newcommand{\net}{\mbox{[Ne \textsc{iii}] }}
\newcommand{\ot}{\mbox{[O \textsc{iii}] }}
\newcommand{\otw}{\mbox{[O \textsc{ii}] }}
\newcommand{\netns}{\mbox{[Ne \textsc{iii}]}}
\newcommand{\otns}{\mbox{[O \textsc{iii}]}}
\newcommand{\otwns}{\mbox{[O \textsc{ii}]}}
\newcommand{\ha}{\mbox{H$\alpha$ }}
\newcommand{\hans}{\mbox{H$\alpha$}}
\newcommand\setItemnumber[1]{\setcounter{enumi}{\numexpr#1-1\relax}}
\usepackage{multirow}

\accepted{\today}

\shorttitle{Star Formation Rates in \nev 3426 \angstrom Selected AGN}
\shortauthors{Feuillet et al.}
\begin{document}

\title{Star Formation Rates in \nev 3426 \angstrom Selected Active Galactic Nuclei: \\ Evidence for a Decrease along the Main Sequence?}

\author[0000-0002-5718-2402]{Léa M. Feuillet}
\affiliation{Institute for Astrophysics and Computational Sciences, Department of Physics, The Catholic University of America,
Washington, DC 20064, USA}

\author[0000-0001-8485-0325]{Marcio Meléndez}
\affiliation{Space Telescope Science Institute, 3700 San Martin Drive Baltimore, MD
21218, USA}

\author[0000-0003-4073-8977]{Steve Kraemer}
\affiliation{Institute for Astrophysics and Computational Sciences, Department of Physics, The Catholic University of America,
Washington, DC 20064, USA}

\author[0000-0003-2450-3246]{Henrique R. Schmitt}
\affiliation{Naval Research Laboratory, Remote Sensing Division, 4555 Overlook Ave SW, Washington, DC 20375, USA}

\author[0000-0002-3365-8875]{Travis C. Fischer}
\affiliation{AURA for ESA, Space Telescope Science Institute, 3700 San Martin Drive, Baltimore, MD 21218, USA}

\author[0000-0002-0982-0561]{James N. Reeves}
\affiliation{Institute for Astrophysics and Computational Sciences, Department of Physics, The Catholic University of America,
Washington, DC 20064, USA}



\begin{abstract}

Studying the behavior along the galaxy main sequence is key in furthering our understanding of the possible connection between AGN activity and star formation. We select a sample of 1215 AGN from the catalog of SDSS galaxy properties from the Portsmouth group by detection of the high-ionization \nev 3426 Å emission line. Our sample extends from $10^{40}$ to $10^{42.5}$ erg/s in \nev luminosity in a redshift range z = 0.17 to 0.57. We compare the specific star formation rates (sSFRs, SFR scaled by galaxy mass) obtained from the corrected \otw and \ha luminosities, and the SED-determined values from Portsmouth. We find that the emission-line-based sSFR values are unreliable for the \nev sample due to the AGN contribution, and proceed with the SED sSFRs for our study of the main sequence. We find evidence for a decrease in sSFR along the main sequence in the \nev sample which is consistent with results from the hard X-ray BAT AGN sample, which extends to lower redshifts than our \nev sample. Although we do not find evidence that the concurrent AGN activity is suppressing star formation, our results are consistent with a lower gas fraction in the host galaxies of the AGN as compared to that of the star forming galaxies. If the evacuation of gas, and therefore suppression of star formation is due to AGN activity, it must have occurred in a previous epoch.

\end{abstract}

%% You will be asked to select these concepts during the submission process
\keywords{Active Galaxies (17) --- Starburst galaxies (1570) --- Star formation (1569)}


\section{Introduction} \label{sec:intro}

% Explanation of the main sequence 
The main sequence (MS; \citealp{Noeske2007}) for star-forming galaxies (SFGs) corresponds to the linear correlation between the star formation rate (SFR) and the mass of the galaxy ($M_*$) in log space. This relationship has been demonstrated extensively and is evident over a range of redshifts, but the correlation varies dramatically between different studies and samples used (see \citealp{Speagle2014} for a review). 

% Deviation from the main sequence/shimizu + negative feedback
Additionally, the introduction of active galactic nuclei (AGN) within the sample modifies the behavior of the relationship (e.g. \citealp{Salim2007, Shimizu2015, Mullaney2015, Stemo2020}). AGN are galaxies containing an accreting supermassive black hole (SMBH) at the center. The presence of this accreting SMBH serves as a powerful source of ionizing radiation which interacts with the ambient gas in the host galaxy. This ionizing radiation is believed to be responsible for the negative feedback process within the AGN (see \citealp{Heckman2014} for a review). Specifically, this radiation would accelerate winds and, in turn, create shocks, which would heat and evacuate the material necessary for star formation from the host galaxy (e.g. \citealp{Silk1998, King2003, DiMatteo2005}. 

The deviation of AGN from the main sequence has been demonstrated by \cite{Shimizu2015} using the \emph{Swift} Burst Alert Telescope (BAT) catalog. The AGN in this sample are hard X-ray selected, which makes it an unambiguous sample of AGN unbiased by the presence of star formation and obscuration. The BAT AGN are local (z $\leq$ 0.05), and they analyze their location with respect to the star-forming MS obtained using far-infrared (FIR) emission from \emph{Herschel} data in the same redshift range. They suggest that this reduction in the SFR is due to AGN feedback which reduces the amount of cold gas that is available for star formation in the AGN. In contrast, our approach consists of using optical data to obtain an unambiguous AGN sample through the use of the \nev 3426 Å emission line. Details regarding our decision to use \nev are given in Section \ref{sec:sample}. Our data also sample a very different redshift range, which corresponds to a different part of cosmological evolution. 

We use optical data from the \emph{Sloan Digital Sky Survey} (SDSS) DR12 galaxy data extracted and made available by the Portsmouth group\footnote{\url{https://www.sdss.org/dr12/spectro/galaxy_portsmouth/}} (Data Release 12; \citealp{Alam2015}). Although the MPA/JHU DR7\footnote{\url{
http://www.mpa-garching.mpg.de/SDSS/DR7/}} sample is much more widely used than the more recent Portsmouth sample (e.g. \citealp{Leslie2016, McPartland2019, Zhuang2019, Torbaniuk2021}), the \nev line is not part of their standard pipeline processing. The SDSS is an optical survey with an extensive galaxy catalog, which allows us to use the \nev optical line to classify our AGN amongst a large initial sample. The Baryon Oscillation Spectroscopic Survey (BOSS) was used to obtain spectral information. Due to the relatively high redshift of the samples used in this paper, the analyzed spectra encompass the entire galaxy, which allows us to take both the host galaxy and the AGN contributions to the emission lines and total spectral energy distributions (SEDs) into account. 

% Outline the paper and the general parameters used throughout the paper
This paper is outlined as follows. Section \ref{sec:sample} details the selection process used to obtain the samples adopted throughout the paper. In Section \ref{sec:main sequence}, we compared three methods of obtaining the SFR for the galaxies in our sample. We then plot our main sequence, which supports the findings of \cite{Shimizu2015} regarding the decrease along the main sequence using a different data set. Finally, we discuss what conclusions can be drawn from the deviation concerning feedback from the AGN to the host galaxy in Section \ref{sec:discussion}. We use \emph{Wilkinson Microwave Anisotropy Probe} (WMAP) 12-year results cosmology throughout, with $H_0$ = 70.0 $km s^{-1} Mpc^{-1}$, and $\Omega _{0}$ = 0.721.

\section{Samples and data selection}\label{sec:sample}

\begin{figure}[ht!]
\centering
\includegraphics[width=0.45\textwidth]{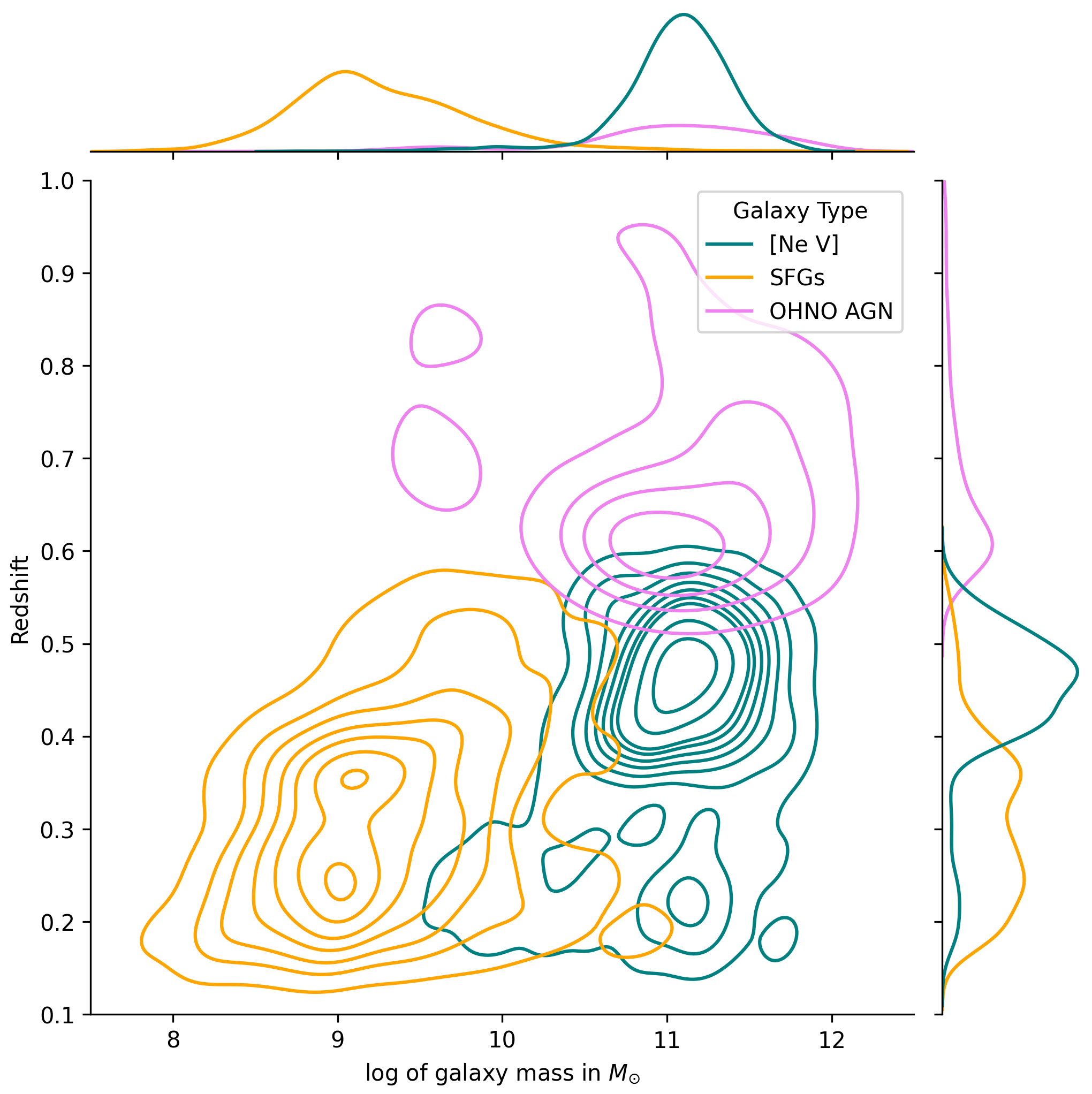}
  \caption{Redshift vs mass distributions for the SFGs, \nev AGN and OHNO classified samples. The redshift was obtained spectroscopically, and the galaxy mass values were calculated using GANDALF \citep{Sarzi2006} by the Portsmouth group.
\label{fig:distributions}}
\end{figure}

\begin{figure}[ht!]
\centering
\includegraphics[width=0.45\textwidth]{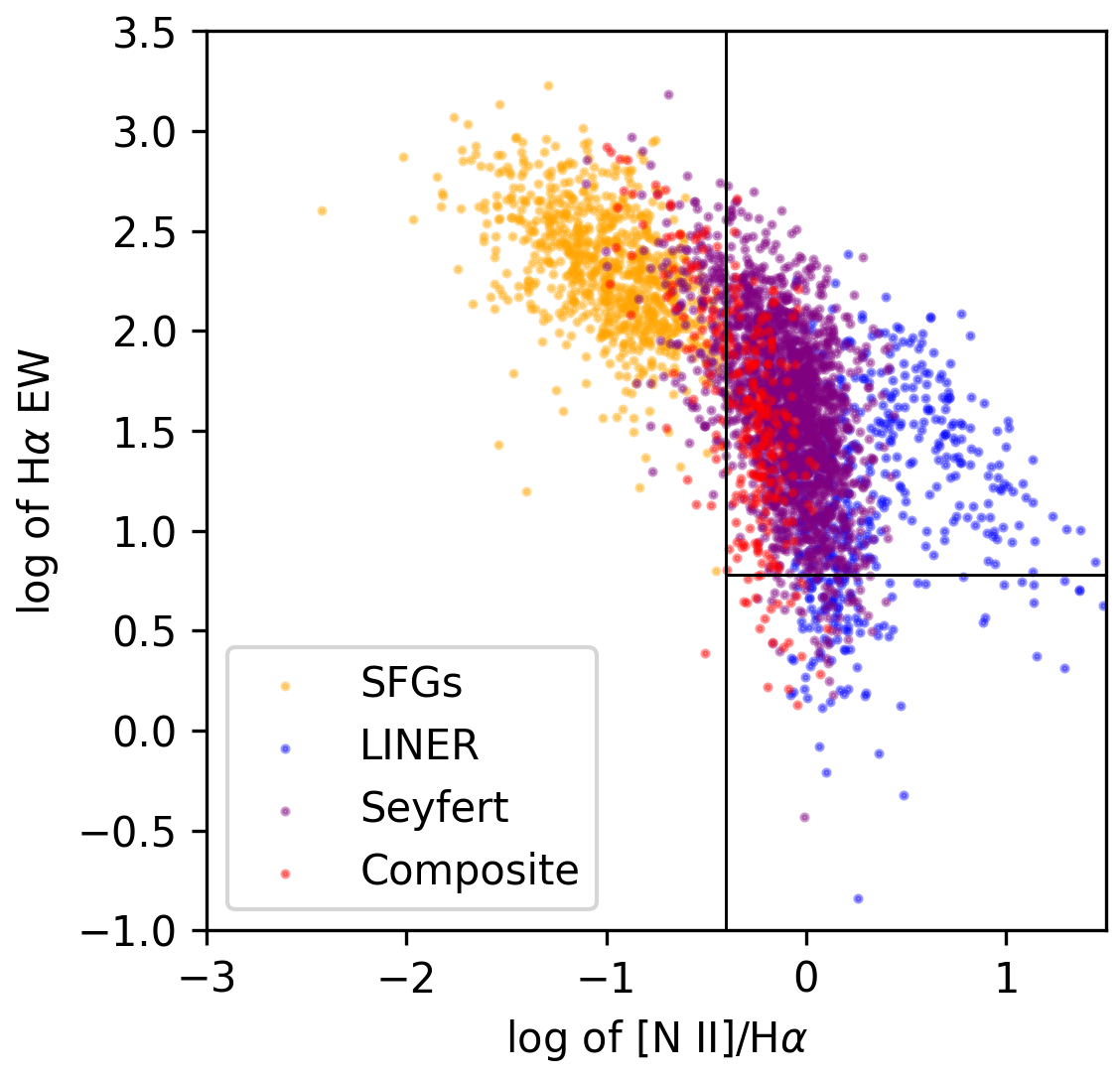}
    \caption{Our \nev AGN and star-forming samples plotted on \cite{Cid2011}'s WHAN diagram, demonstrating the reliability of using \nev to select AGN, as 90\% are indeed in the AGN region. 
    \label{fig: whan}}
\end{figure}

The sample and data selection process is explained in detail in \cite{Feuillet2024}. To summarize, the data in the Portsmouth group's emissionLinesPort table is used to obtain emission-line equivalent widths (EWs), fluxes and their uncertainties, and redshifts for the galaxies \citep{Thomas2013}. The emission-line fluxes were corrected for interstellar reddening by using the observed and theoretical Balmer decrements (\mbox{H$\alpha$/H$\beta$}) and the \cite{CCM1989} reddening law for $R_{v}$ = 3.1, to obtain the E(B-V) values. We also require an S/N $\geq$ 3 for the \mbox{\net 3869 Å}, \mbox{\ot 5007 Å}, \mbox{\otw 3726, 3728 Å}, and H$\alpha$ emission-lines (hereafter \netns, \otns, and \otwns). We then separate the galaxies into two samples: the SFGs sample and the full AGN sample, which have been classified using the standard BPT diagram. The \nev sample is a sub-sample of the full AGN sample, corresponding to all AGN with a S/N $>$ 3 for the \nev emission-line flux.

Although the \ot 5007 \angstrom has been preferred over the \nev 3426 \angstrom line for AGN classification purposes, \nev allows for a greater redshift range (up to \mbox{z $\leq$ 2}) when using SDSS and BOSS, despite being weaker. The use of \nev for this purpose has been demonstrated in both the IR (see \citealp{Abel2008, Satyapal2008, Goulding2009}) and the optical (e.g. \cite{Schmidt1998}, \cite{Gilli2010}, and \cite{Mignoli2013}). It has been shown by \cite{BPT1981}, that none of the HII regions investigated showed any amount of \nev within their spectra, as even the hottest main sequence stars do not emit photons energetic enough to produce \nevns. Ne$^{+4}$ has a high ionization potential \mbox{(97 eV)}, can only be produced by high-energy phenomena such as shocks and the presence of AGN, and is thus an ideal AGN tracer. 

In this paper, we make additional use of the stellarMassStarformingPort table which contains the star-formation rates and galaxy masses. Both are calculated by the Portsmouth group through SED fitting, similar to the method used in \cite{Shimizu2015}, with models found in \cite{Maraston2006}. We direct readers to \cite{Maraston2013} for further information on the determination of stellar masses.We use the values obtained using a \cite{Salpeter1955} initial mass function (IMF) and convert them to a Chabrier IMF using a 0.67 factor throughout the present paper, as suggested in \cite{Madau2014}.

The redshift for the \nev sample extends from \mbox{0.168 $\leq$ z $\leq$ 0.568,} with galaxy masses ranging within \mbox{9.150 $\leq$ log $M_*$ $\leq$ 11.890}. To compare galaxies at similar redshifts, we limit the SFGs to the same range as the \nev AGN. For comparison with higher redshift galaxies, we use the OHNO diagram presented in \cite{Feuillet2024} to get a sample of AGN that extends to a redshift of 1.07. The full distributions for the redshift and mass of the SFGs, \nev AGN, and OHNO AGN samples are given in Figure \ref{fig:distributions}. We also plot our SFGs and \nev AGN on the WHAN diagram \citep{Cid2011} (see Figure \ref{fig: whan}). The WHAN diagram plots the \ha EW against [N \textsc{ii}]/\ha, with delimitation lines separating the SFGs from the AGN and Seyfert galaxies from LINERs. 90\% of the \nev AGN sample also falls in the AGN region of the WHAN diagram. We will now use our SFGs and AGN samples to investigate the MS.  

\section{Main Sequence}\label{sec:main sequence}

\subsection{Star Formation Rates}\label{sec:SFR}

Determining the SF galaxy MS is dependent on the availability of accurate SFRs, which can be obtained using a variety of methods (see \citealp{Kennicutt2012} for a review). However, the presence of an AGN can influence the resulting SFR values. In the case of SED fitting, the AGN affects the shape of the SED and the light from the AGN can be interpreted as coming from the formation of young stars, thereby increasing the resulting SFR \citep{Pacifici2023}. \cite{Ciesla2015} found that by not omitting an AGN component when using SED fitting to obtain the SFRs of AGN, as is the case for our SED-determined SFRs, the ratios are overestimated\footnote{\cite{Ciesla2015} also found that the resulting stellar mass estimates were not affected by the inclusion or omission of an AGN component when it comes to type 2 AGN.}.  

When looking at emission-line derived SFR, the results depend on the contribution of the AGN to the total flux observed. Various studies have determined contributions of the AGN to the fluxes of the emission lines used as star formation tracers. For instance, \cite{Melendez2008} investigated a sample of nearby Seyfert galaxies observed with \emph{Spitzer}'s Infrared Spectrogram (IRS). They found that 34-75\% of the flux of the IR line \mbox{[Ne \textsc{ii}]} 12.81 $\mu$m, which has been used as a SFR indicator similar to \otw \citep{Cao2008}, is due to the presence of the AGN (see also \citealp{Melendez2014}). \cite{Thomas2018} also report contributions to the \otw flux ranging between 20\% and 100\%, while \cite{Davies2014} found contributions of about 40\%. The additional contribution from the AGN thus leads once again to an overestimation of the SFR values. 

Evidently, according to the literature, neither SED nor emission-line determined values may give perfect estimates of the SFRs for our AGN. We now compare the different SFR values available for our sample to investigate the biases in further detail. The Portsmouth group provides values for both the SFR and mass by fitting the galaxies' SED using the method described in \cite{Maraston2013}. The SDSS spectrum range contains the \otw and H$\alpha$ emission lines, which we may also use to obtain SFR values.

\cite{Kewley2004} use a sample of purely SFGs, excluding any potential AGN, to derive their \otwns-based SFR equation. The relationship specifically requires the luminosity values to be corrected for reddening. The equation is given as follows:
\begin{equation}\label{eq: kewley}
    \textrm{SFR([O II])} = (6.58 \pm 1.65) * 10^{-42}  \textrm{ L([O II])}
\end{equation}
in which SFR(\otwns) is the star formation rate, and L(\otwns) is the luminosity of the \otw emission line \citep{Kewley2004}. Additionally, the \hans-based SFR is obtained using:
\begin{equation}\label{eq: kewley}
    \textrm{SFR(\hans)} = 7.9 * 10^{-42}  \textrm{ L(\hans)}
\end{equation}
in which SFR(\hans) is the star formation rate, and L(\hans) is the luminosity of the \ha emission line \citep{Kennicutt1998}. Using these relationships and the masses obtained by the Portsmouth group, we also define the specific star-formation rate (sSFR), which corresponds to the SFR of a galaxy divided by its total mass. 

\begin{figure}
  \centering
    \includegraphics[width=0.45\textwidth]{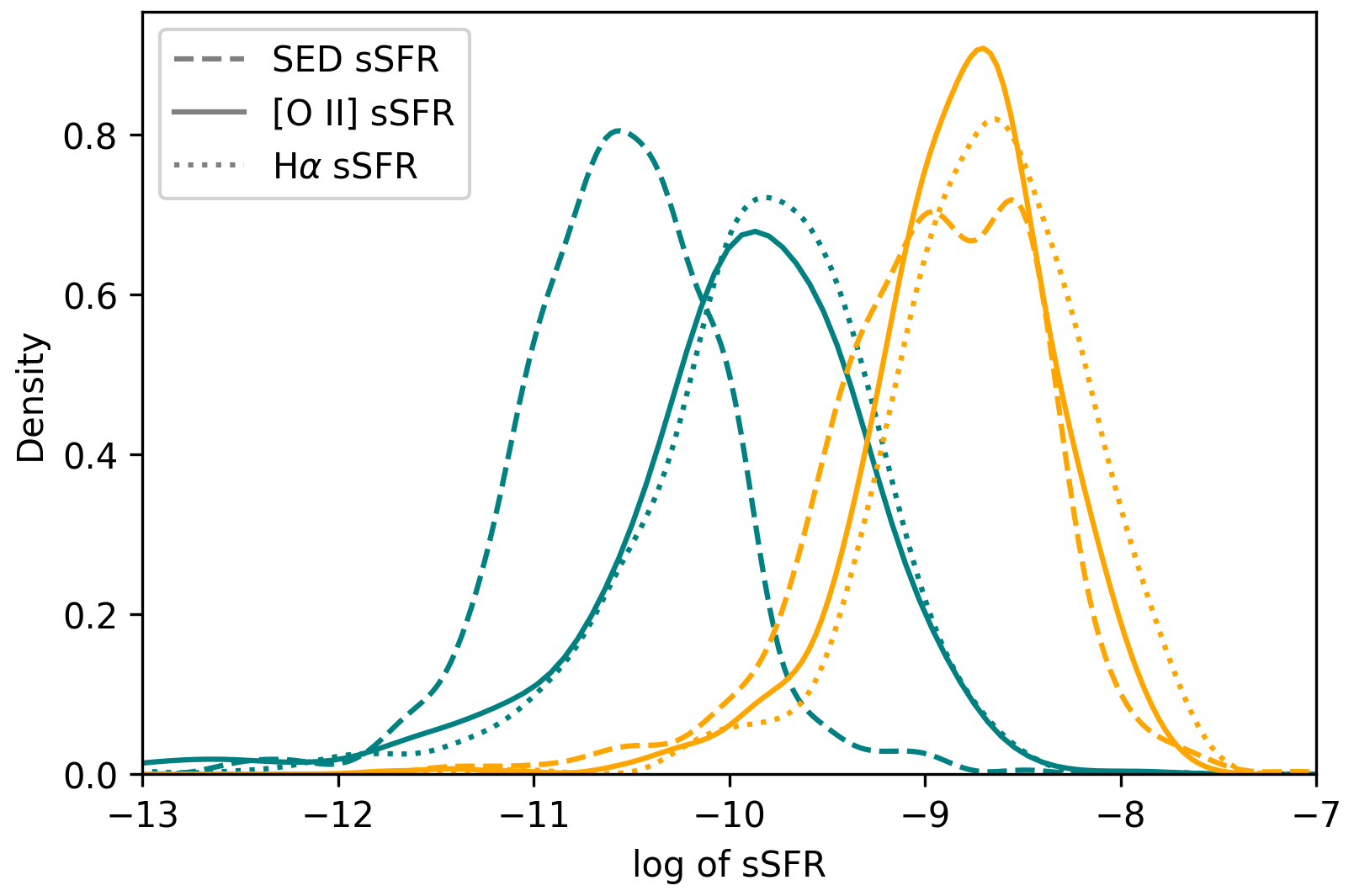}
\caption{Comparison of the sSFR values from Portsmouth's SED fitting method to the \otw emission-line equation from \cite{Kewley2004} and the \ha relationship from \cite{Kennicutt1998}. We show the distribution of the sSFR values for the \nev AGN (green) and SFG (orange) samples. The \otw and \hans-based values are higher than the SED value because of the contribution to both emission lines by the AGN.}
\label{fig:sSFR dist}
  \end{figure}

We obtained sSFR values for our SFGs and \nev AGN samples and plotted the estimated probability density for the results of all three methods (see Figure \ref{fig:sSFR dist}). Looking at the star-forming sSFR distributions, we can see that all three overlap considerably. This suggests that all three methods give similar results when applied to the SFGs. On the other hand, the \nev AGN galaxies show a significant deviation between the SED and the \ha and \otw methods, with the former showing overall lower sSFR values compared to the other two. This is likely due to the aforementioned contribution of the AGN to the emission lines, which does not appear to affect the SED fitting method as significantly. To mitigate the overestimation of the SFR, we proceed with the use the SED SFR and sSFR values to investigate the MS for our samples, with the caveat that there still may be inherent uncertainties due to the presence of the AGN.

\subsection{Behavior Along Our Main Sequence}

\begin{figure*}
  \centering
    \includegraphics[width=\textwidth]{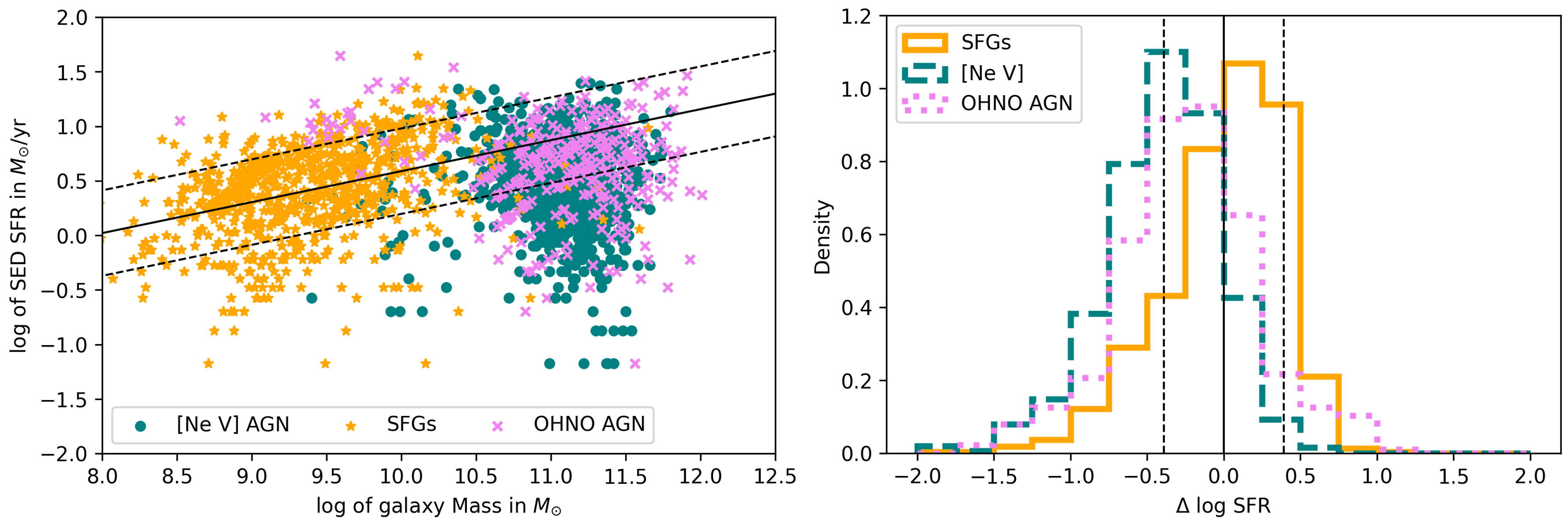}
\caption{Left panel: Our three samples plotted on the main sequence. The solid black line corresponds to the best fit main sequence for the SFGs, and the dashed ones show the $\pm$1$\sigma$ lines. We can see the \nev AGN falling below the main sequence. Right panel: Distributions showing the distance from the main sequence explicitly for all three samples.}
\label{fig:SFR_MS}
  \end{figure*}

Using the SED-determined SFRs and masses provided by the Portsmouth group, we plot the main sequence for our SFGs. We fit the SFGs using linear regression and find that the equations for our MS is: 
\begin{equation}
    \textrm{log SED SFR} = 0.28 \textrm{ log } M_* - 2.25
\end{equation}
We also calculate the standard deviation ($\sigma$) for the regression lines\footnote{The standard deviation for the regression line is obtained using the following equation: $S^{}_{y/x}=\sqrt{S_{r}^{}/(n-2)}$, where $S_{r}^{} = \sum_{i=1}^{n}(y^{}_{i}-a_{0}^{}-a^{}_{1}x^{}_{i})^{2}$ is the sum of squares of the residuals with respect to the regression line, $a_{0}^{}$ and $a_{1}^{}$ are the intercept and slope of the regression line respectively, and n is the number of data points.}, which is $\sigma_{SED/\otwns}$ = 0.39. The MS regression lines and the $\pm$1$\sigma$ lines are all plotted on top of the MS in Figure \ref{fig:SFR_MS}, left panel. We have also plotted the \nev AGN sample on the MS plot, with the deviation already apparent. While the majority of the SFGs are included within 1$\sigma$ of the MS, the \nev AGN have a significant number falling below in both plots.

We now study the full SFG and \nev AGN samples in more detail, and adopt a similar convention to study the deviation from the main sequence as those of \cite{Shimizu2015}. We calculate the distance from our MS as such:
\begin{equation}
    \Delta \textrm{ log } M_* = \textrm{log SFR}_{SED} - \textrm{log SFR}_{MS}
\end{equation}
where log SFR$_{SED}$ is the SED-determined SFR for each galaxy and log SFR$_{MS}$ is the theoretical SFR given by the equation of our MS. The distributions for the resulting values are plotted on the left panel of Figure \ref{fig:SFR_MS}. 

\begin{table}[h!]
\centering
\caption{Summary of the percentages of galaxies below, within, and above the MS defined by the SFGs. }
\label{tab:MSvals}
\begin{tabular}{llll}
\hline
Sample     & $\leq$ -1$\sigma$ & in MS &  1$\sigma$ $\leq$ \\
\hline
SFGs & 15\% & 72\%  & 13\%  \\
\nevns & 49\%  & 50\%  & 1\%  \\
OHNO  & 32\%    & 61\%  & 7\%  \\
\hline
\end{tabular}
\end{table}

The \nev AGN having approximately half of the population lying below the MS using both SFR options is significant and confirms the deviation from the MS that could already be seen on the scatter plot (see Table \ref{tab:MSvals} for the full description of the values). Once again, as the SED-derived SFRs are likely to be overestimates, it is probable that the true deviation from the MS is even greater than seen in Figure \ref{fig:SFR_MS}. The OHNO AGN have a similar distribution to the \nev AGN when using the SED-determined SFRs, with 32\% below the MS, but represent a higher redshift range. This suggests that the phenomenon is present at least up to a redshift of 1.06.

\subsection{Redshift and Mass Dependence of the Deviation from the MS}

\begin{figure}
  \centering
    \includegraphics[width=0.45\textwidth]{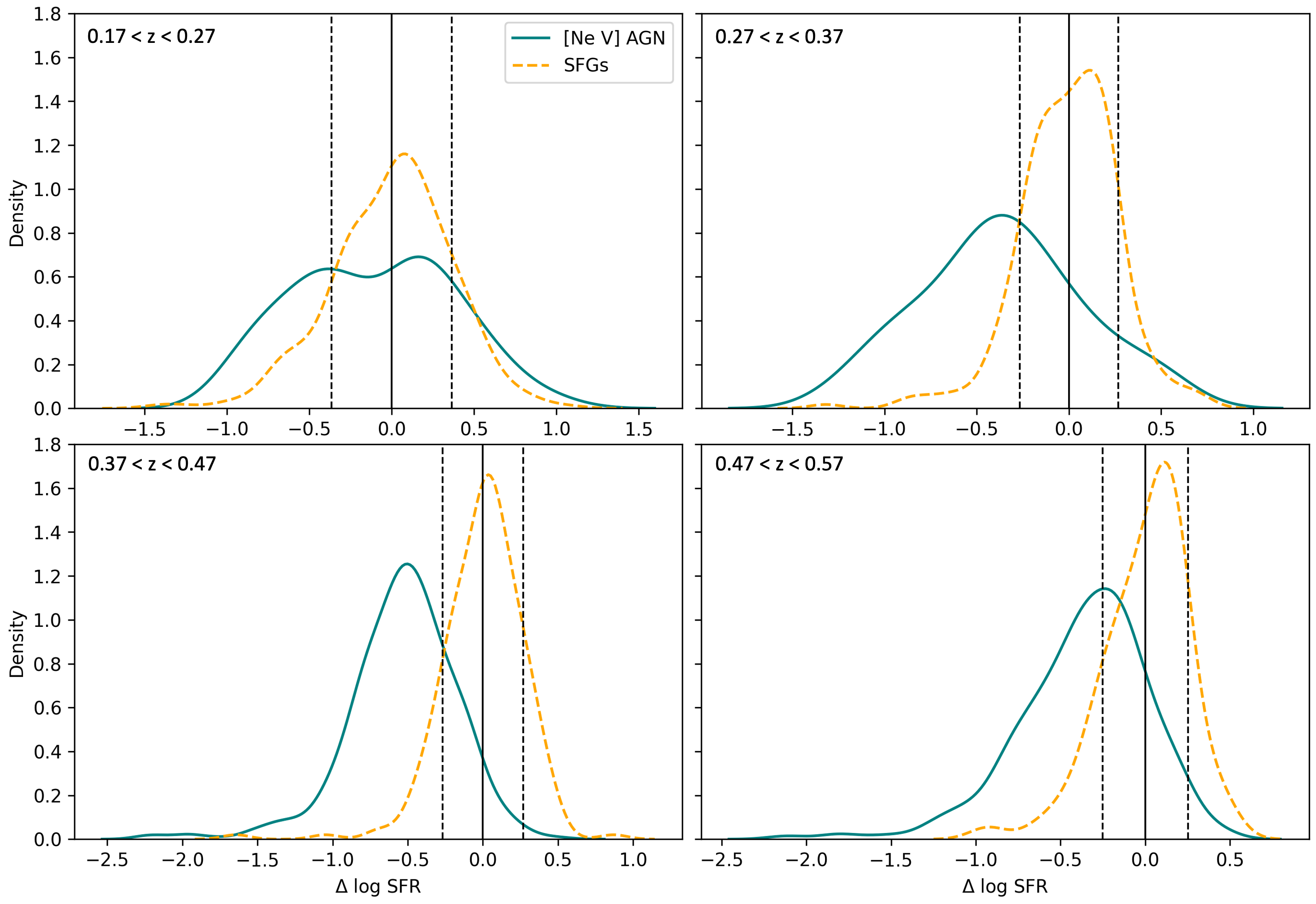}
\caption{Redshift dependent plots showing the deviation from the MS. The MS as well as the \nev AGN are plotted in ranges of 0.1 in redshift, from 0.17 to 0.57. The deviation from the main sequence is clear in all plots, with the caveat of the small \nev sample size in the first. This demonstrates that there is no clear redshift dependence.}
\label{fig:SFR_MS_z}
  \end{figure}

It is well known that the SFR has varied since the beginning of the Universe, with a peak in star formation density occurring around z = 2 \citep{Madau2014}. While the redshift ranges of the samples overlap as seen in Figure \ref{fig:distributions}, we can also see that the \nev AGNs are concentrated at a higher redshift compared to the SFG sample. The AGN are therefore closer to the peak in SF, and would therefore tend toward having higher SFRs, which is the opposite of what we have shown in this paper. Additionally, further investigation of the redshift-dependence of the deviation from the MS shows that it is apparent regardless of redshift. As can be seen in Figure \ref{fig:SFR_MS_z}, the deviation is very clear at all redshift ranges, with the exception of the first where the ambiguity stems from the small \nev AGN sample size. Hence, we are confident that the deviation from the main sequence is truly the result of AGN activity, as also suggested by \cite{Shimizu2015}. 

Due to the use of the \nev emission-line to obtain an unambiguous sample of AGN, our resulting \nev AGN tend towards having higher masses, as seen in Figure \ref{fig:distributions}. Indeed, the galaxies must be powerful enough to produce a significant amount of \nev flux for the S/N ratio to be greater than 3, which requires a luminous AGN. AGN luminosity is a function of both the accretion rate and the black hole mass. However, even AGN with sub-Eddington accretion rates such as Mrk 3 \citep{COllins2009}, show strong \nev provided that they SMBH of sufficient mass. e.g., $>$ 10$^7$ M$_\odot$. Therefore, given the relationship between bulge mass and the mass of the central SMBH \citep{Ferrarese2002, Bandara2009}, strong \nev would preferentially occur in the higher mass galaxies. Unfortunately, this introduces a dichotomy between the masses of the SFGs and the \nev AGN. This implies that it is statistically impossible to differentiate between the deviation from the main sequence being applicable to high mass galaxies or to AGN specifically. 

\begin{figure}
  \centering
    \includegraphics[width=0.45\textwidth]{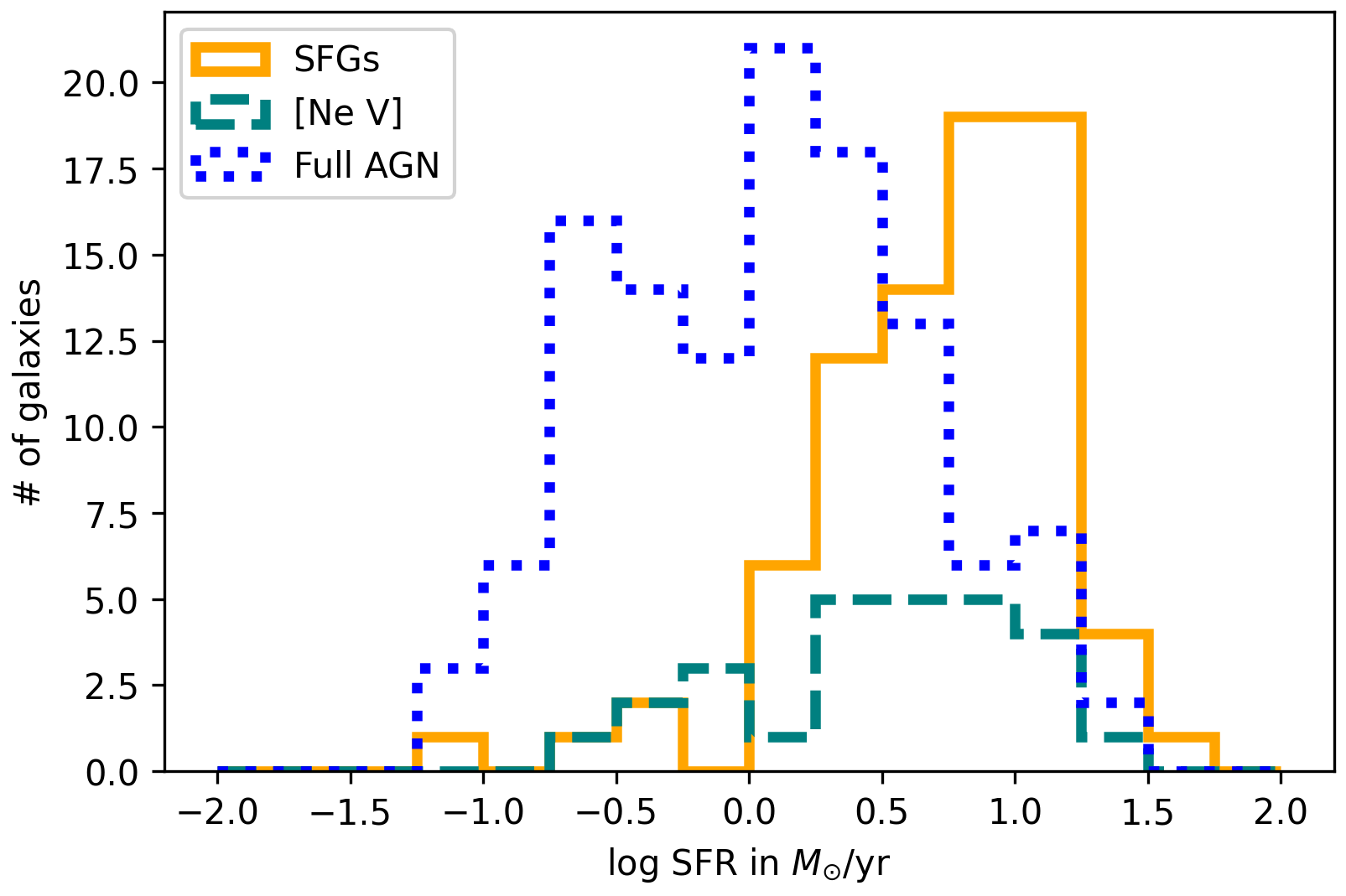}
\caption{This plot shows the distribution of the SFRs for the SFGs, the \nev AGN, and the full AGN sample within the restricted mass range of log M$_{*}$ = 10-10.5 (see Section \ref{sec:sample} for sample descriptions). This shows once again that the AGN tend towards having lower SFRs compared to the SFGs, even within a fixed mass range.}
\label{fig:SFR_MS_fixed_mass}
  \end{figure}

The lack of overlap in mass means that we can only investigate the deviation from the main sequence at fixed mass for one range, between log M$_{*}$ = 10-10.5 for both our \nev AGN, and the full AGN sample. Performing a partial correlation test on this full AGN sub-sample, we find that there is a very low correlation between the SFR and the mass of the galaxy (r = 0.02), while there is a high correlation between the SFR and the galaxy type (r = 0.36). We also plot the distributions of the SFRs in Figure \ref{fig:SFR_MS_fixed_mass}, which clearly shows the full AGN sample falling below the SFGs, as well as the \nev AGN to a smaller extent. 

\subsection{Comparison With Previous Main Sequences}

\begin{figure*}
  \centering
    \includegraphics[width=\textwidth]{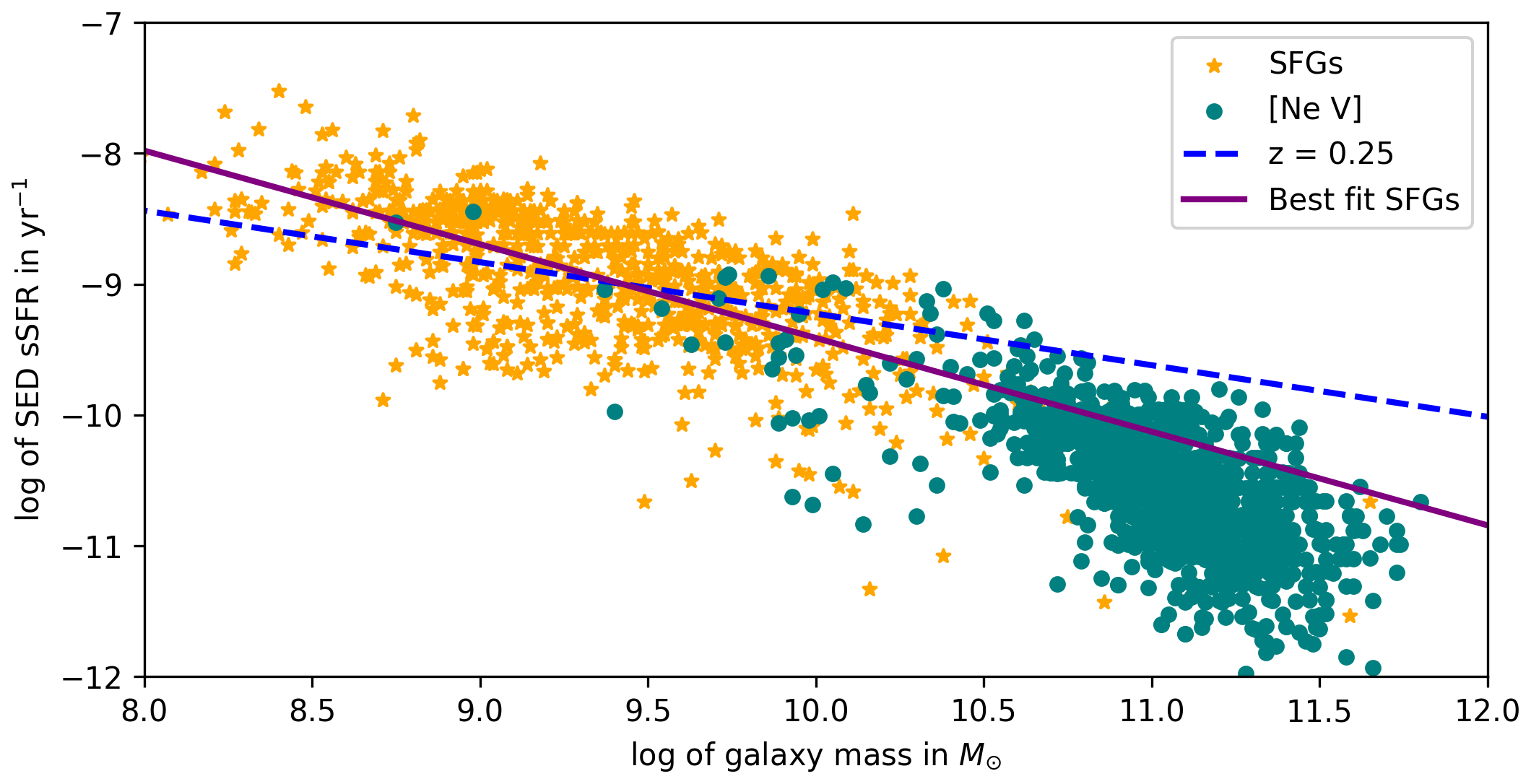}
\caption{Comparing our sSFR MS (purple, solid) for our SFGs and [Ne V] AGN, with the relationship found in \cite{Speagle2014} for z = 0.25 (blue, dashed). We find that our SFG data fits well with the \cite{Speagle2014} relationships, which span a similar range in redshift as our SFGs. However, the \nev AGN have a much steeper slope than the \cite{Speagle2014} lines, indicating once again a deviation from the MS.}
\label{fig:main seq}
  \end{figure*}

In addition to the standard MS plotting the SFR against the galaxy mass, we also investigate the sSFR MS and compare our distribution to previously determined main sequence regression lines. In Figure \ref{fig:main seq}, we compare relationships found in \cite{Speagle2014} and \cite{Shimizu2015} to our SFGs and \nev AGN data. These relationships have been calculated to specifically fit SFGs, without the influence of AGN. 

The \cite{Speagle2014} relationship was obtained by combining a large number of MS fits found in the literature at a range of redshifts. The resulting relationship is thus time-dependent and is defined as: 
\begin{equation} \label{eq: speagle}
    \textrm{log SFR}(M_*, t) = (0.96 - 0.045t) \textrm{ log } M_*  - (7.41 - 0.27t)
\end{equation}
where SFR(M$_*$, t) is the star formation rate as a function of $M_*$, the mass of the galaxy in units of solar masses ($M_\odot$), and the time t, which corresponds to the age of the universe in Gyr. For our analysis, we further convert the time dependence in Equation \ref{eq: speagle} to a redshift one using the following equation:
\begin{equation} \label{eq: time to redshift}
    t(z) = \frac{2}{3 H_0 \Omega_0^{1/2} (1+z)^{3/2}}
\end{equation}
where t is again the age of the universe in Gyr, z is the redshift, \mbox{$H_0$ = 70.0 $km s^{-1} Mpc^{-1}$}, and \mbox{$\Omega _{0}$ = 0.721} as previously stated. The slope and intercept of the main sequence both go down as t increases or as z decreases. This is the case as the SFR is higher in younger galaxies. In Figure \ref{fig:main seq}, we plot the relationship at a redshift of z = 0.25 which is close to the average redshift of our samples. 

The \cite{Shimizu2015} relationship is, however, only dependent on galaxy mass as they used exclusively local (z $\leq$ 0.05) galaxies. It is calculated as the best-fit line to \emph{Herschel} Reference Survey and \emph{Herschel} Stripe 82 survey data:
\begin{equation} \label{eq: shimizu}
      \textrm{log SFR}(M_*) = 1.01 * \textrm{log } M_* - 9.87  
\end{equation}
where the variables are the same as in Equation \ref{eq: speagle}. As implicitly shown in this equation, the slope of the sSFR MS relationship from the \cite{Shimizu2015} paper will be flat. It is clear by looking at Figure \ref{fig:main seq} that a flat relationship does not fit with our data. This is perhaps due to the inclusion of lower mass galaxies in the Herschel sample and possible biases in the BAT sample being hard X-ray selected. 

However, as seen in Figure \ref{fig:main seq}, our specific star formation main sequence plot trends are in overall agreement with the redshift-dependent relationship of \cite{Speagle2014}, having a similar downward trending slope\footnote{The Speagle slope fits more closely with our MS when including a similar mass range as in their paper and removing outliers.}. We can also see that there is a noticeable decrease in the sSFR with increasing mass in the \nev sample. This trend is not as pronounced in the SFG sample, which suggests that the trend is affected by the presence of the AGN. 

Our analysis is complementary to BAT sample analysis by \cite{Shimizu2015}, as we both use a specific criterion to unambiguously target AGN. Our results regarding the AGN deviation from the main sequence support the ones found by \cite{Shimizu2015} despite the lack of overlap between our galaxies and the BAT sample, with the BAT AGN having a maximum redshift of z = 0.05. Being able to use the optical \nev line as opposed to X-ray-selected AGN makes the result more accessible since it increases the possibility of detecting fainter objects.

\section{Discussion}\label{sec:discussion}

In the previous section, we found that a significant percentage of the \nev AGN fall below the star-forming main sequence, indicating a decrease in SFR at higher masses. This deviation from the main sequence has been used as an indicator of feedback in AGN, with the AGN causing the suppression of star formation via various processes. In this section, we further investigate this possibility by looking at the bolometric luminosity of our \nev AGN.

We calculated the bolometric luminosity (L$_{bol}$) for the galaxies in the \nev sample using both the \ot and the \nev luminosities (L$_{[O III]}$ and L$_{[Ne V]}$). We used the \ot bolometric correction for reddening corrected fluxes from \cite{Lamastra2009}, which corresponds to multiplying L$_{[O III]}$ by a factor of 454. Although different factors are suggested depending on the L$_{[O III]}$ range, we decided to apply a single correction factor instead of differentiating between the AGN above and below the log L$_{[O III]}$ = 42 threshold, as doing so created an artificial gap in the resulting plot. Additionally, 87\% of the sample lies above the log L$_{[O III]}$ = 42 limit.

We also used the relationship from \cite{Satyapal2007}:
\begin{equation}
    \textrm{log } L_{bol} = 0.94* \textrm{log } L_{[Ne V]_{IR }}+6.32
\end{equation}
where L$_{bol}$ is the bolometric luminosity, and L$_{[Ne V]_{IR}}$ is the luminosity of the IR \nev 14.32 $\mu$m line. We used a \nev 14.32 $\mu$m/\nev 3426 \angstrom ratio of 1.60 to calculate the bolometric luminosity using the optical \nev line instead. The ratio was obtained using the Cloudy photoionization modeling code (v. 17.0, \citealp{Cloudy2017}). The parameters used are the same as those outlined in \cite{Feuillet2024}. However, we use a log U = -1.1 to -1.3 and log $n_{H}$ = 3 cm$^{-3}$ and averaging the results. We used a higher ionization parameter than in the previous models as we set out to optimize the parameters for \nev emission \citep{Ferguson1997}. The two $L_{bol}$ values follow a similar trend with an offset of about half an order of magnitude and have a general overlap.

\begin{figure}[ht!]
\centering
\includegraphics[width=0.47\textwidth]{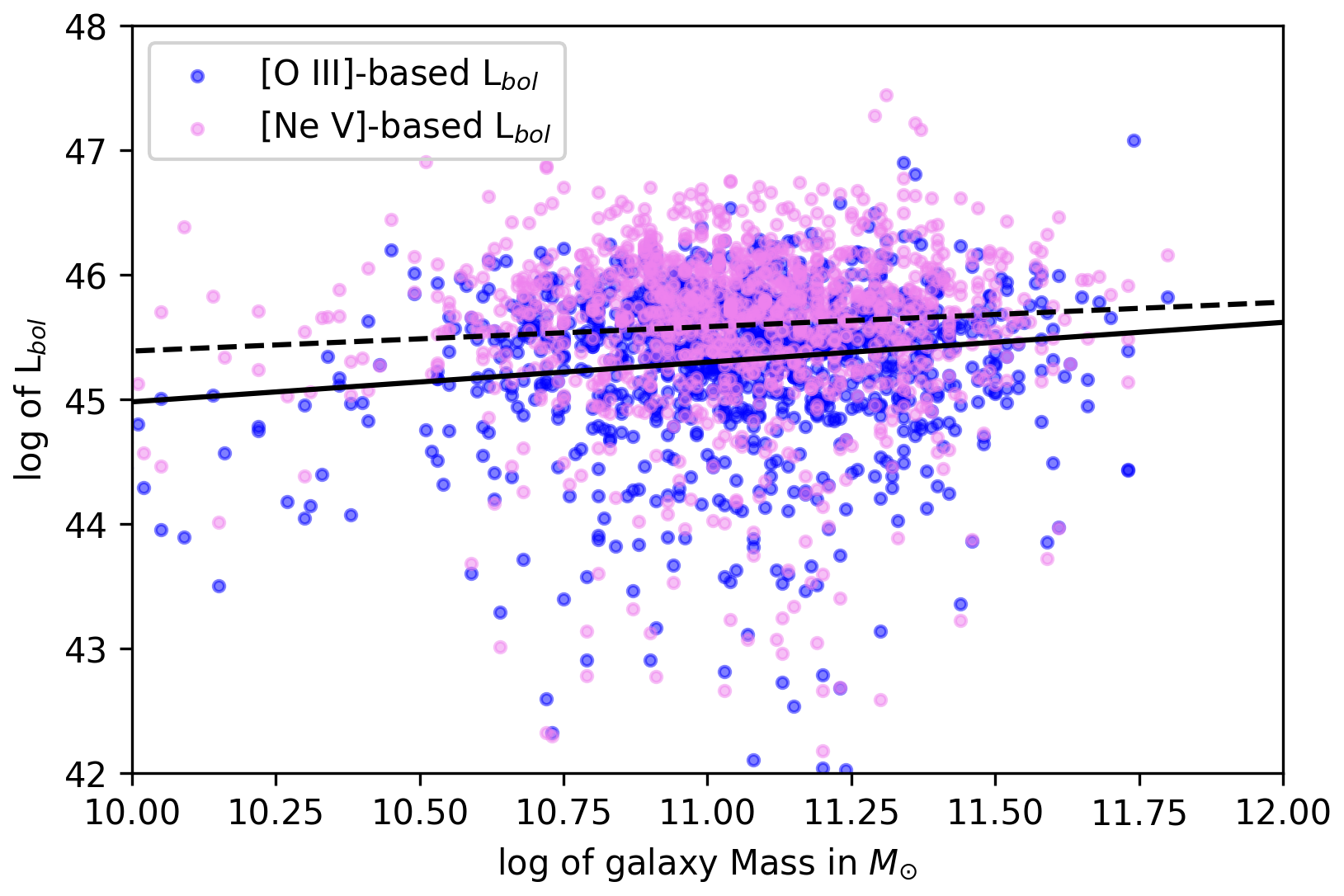}
    \caption{Bolometric luminosities for the \nev sample obtained using two methods reliant on \ot and \nev respectively. The \otns-based bolometric luminosity is calculated using the 454 factor from \cite{Lamastra2009}. The \nev bolometric luminosity uses both the relationship between L$_{bol}$ and the luminosity of the IR \nev 14.32 $\mu$m line from \cite{Satyapal2007}, as well as a conversion factor between the IR and optical \nev lines extracted from the previously used NLR Cloudy models. The solid and dashed black lines represent the line of best fit of the \nevns- and \otns-based bolometric luminosities respectively.
    \label{fig:bol}}
\end{figure}

Looking at the resulting plot containing both the \ot and \nev based bolometric luminosities, we see that neither appears to increase monotonically in any significant way (see Figure \ref{fig:bol}). Such an increase would have been expected as the mass of the galaxy is proportional to the mass of the central SMBH \citep{Ferrarese2002, Bandara2009}. A higher mass SMBH would also be related to an increase in the amount of radiation for a given accretion rate, and thus increase the $L_{bol}$ of the AGN. This result leaves us with two options. The first assumes that \ot is a good measure of bolometric luminosity, which would indicate that the decrease in sSFR does not have anything to do with the strength of the AGN, as we would have otherwise expected $L_{bol}$ to increase with the mass of the galaxy.

The second option would be that \ot is not a good measure of bolometric luminosity. One possibility is that it could be due to over-ionization \citep{Baldwin1977}. As \ot is a measure of the ionizing radiation if over-ionized the resulting gas would be transparent and would not be able to re-radiate the entirety of the ionizing radiation. It would theoretically be possible for the \nev to also be over-ionized, but is unlikely given its high ionization potential. Therefore, the lack of divergence between the \ot and \nev determined bolometric luminosities suggests that over-ionization is not the most likely explanation.

The other reason would be a lack of gas in the galaxy. For example, the galaxy could have gone through a blowout stage, during which AGN feedback, or some other mechanism (e.g. \citealp{Heckman2014}), ejects gas from the host galaxy. Assuming that a majority of the AGN in our sample are past this blowout stage, this would explain both the decrease in sSFR and "specific bolometric luminosity". Indeed, there would not be enough gas to continue forming stars nor would there be as much left to ionize into \otns. In agreement with other papers, our results do not suggest current AGN activity being responsible for the decrease in the SFR, but rather residual or reactivated AGN activity after the feedback has already ceased \citep{Peng2010, Mulcahey2022, Harrison2023}. The fact that we can see the deviation from the main sequence for the BAT, \nev, and OHNO AGN, spanning redshifts up to 1.06 indicates that the blowout phase must have occurred before then. 

Another indication of this explanation arose when we looked at the SFR distributions for our samples and compared them to those of \cite{Shimizu2015} in Figure \ref{fig:SFR_MS}. They each represent a different redshift span, with the BAT sample being local (z $\leq$ 0.05), our SFGs and \nev AGN ranging between 0.168 $\leq$ z $\leq$ 0.568, and the OHNO AGN having redshifts from 0.568 to 1.07. The local MS showed that the SFR for the SFGs range between $\approx$ -1.5 to 0.5 log SFR, while the BAT AGN span from $\approx$ -1 to 1.5. Our SFGs have an SFR range that is overall 1 dex higher than the local galaxies. This can be explained by the fact that the galaxies in our sample live closer to the peak in SFR, which occurred between 2 $<$ z $<$ 3, with a sharp decline of about 1 dex from \mbox{z = 1} and \mbox{z = 0} \citep{Hopkins2006}. 

However, the BAT AGN, our \nevns-selected AGN, and the OHNO AGN all appear to have almost identical ranges of SFR, despite representing different redshift ranges. The process responsible for the reactivation of the AGN would be also fueling the star formation within the galaxy, keeping the SFR constant over the range in redshift. 

\section{Conclusion} \label{sec:cite}

By applying several criteria to the full galaxy sample, we separated it into a SFG and a \nev AGN sample in order to investigate their differences in behavior along the galaxy main sequence.

\begin{enumerate}
    \setItemnumber{1}
    \item We calculated the specific SFR using three different methods: SED fitting provided by the Portsmouth group, \otwns-based sSFR using the relationship in \cite{Kewley2004}, and \hans-based from \cite{Kennicutt1998}. While all three methods gave consistent results for the SFG sample, the emission-line methods overestimated the SFRs of the AGN sample compared to the SED fitting values. Despite the limitations of both methods, we opted to use SED fitting values for both samples.
    \setItemnumber{2} 
    \item We plotted the MS for our SFG and \nev AGN samples, and found that there is a deviation when it comes to the AGN. While only 15\% of the SFGs fall below the MS, 49\% of the \nev AGN sample lies under, which indicates a reduction in SFR in almost half of the galaxies. This result is in agreement with that of \cite{Shimizu2015}, which used the BAT AGN sample. 
    \setItemnumber{4}
    \item We also plotted the specific SFR MS and compared our data to previously determined MS relationships. Our data agrees well with the work of \cite{Speagle2014}. We found a rapid decrease in sSFR in the AGN that is not as pronounced in SFGs, again suggesting a decrease in SFR with increasing mass in AGN.
    \setItemnumber{5}
    \item We looked at the bolometric luminosity of our \nev AGN sample using two different methods and found that the bolometric luminosity does not have a mass dependence. One possible explanation is that the deviation is due to a reduction in the SF in the AGN as a result of past activity. However, we cannot reach a definite conclusion with this data set.
\end{enumerate} 

%% Also note that the akcnowlodgment environment does not support long amounts of text. If you have a lot of people and institutions to acknowledge, do not use this command. Instead, create a new \section{Acknowledgments}.

\begin{acknowledgments}
This work has made use of SDSS DR12 data. Funding for the Sloan Digital Sky Survey IV has been provided by the 
Alfred P. Sloan Foundation, the U.S. Department of Energy Office of Science, and the Participating Institutions. SDSS-IV acknowledges support and resources from the Center for High Performance Computing  at the University of Utah. The SDSS website is www.sdss4.org. The authors would like to thank the Portsmouth Group for making their data tables available online. The data can be found at \url{https://www.sdss4.org/dr12/spectro/galaxy_portsmouth/}. 
\end{acknowledgments}






\bibliography{Feuillet-Lea}{}
\bibliographystyle{aasjournal}



\end{document}